\documentstyle[12pt,aasms4]{article}
\lefthead{Oka et al.}
\righthead{A HIGH VELOCITY MOLECULAR CLOUD NEAR THE CENTER OF THE GALAXY}

\def\kms{\hbox{km s$^{-1}$}}
\def\VLSR{\hbox{$V_{\rm LSR}$}}

\begin{document}
\title{A HIGH VELOCITY MOLECULAR CLOUD NEAR THE CENTER OF THE GALAXY}
\author{Tomoharu Oka\altaffilmark{1}, Glenn J. White\altaffilmark{2}, Tetsuo Hasegawa\altaffilmark{3}, 
Fumio Sato\altaffilmark{4}, Masato Tsuboi\altaffilmark{5}, 
and Atsushi Miyazaki\altaffilmark{5}}

\altaffiltext{1}{{\footnotesize The Institute of Physical and Chemical Research (RIKEN), 
2-1 Hirosawa, Wako, Saitama 351-0198, Japan.}} 
\authoremail{oka@postman.riken.go.jp}
\altaffiltext{2}{{\footnotesize Physics Department, Queen Mary \& Westfield College, University of London, 
Mile End Road, London E1 4NS, U.K.}} 
\altaffiltext{3}{{\footnotesize Institute of Astronomy, Faculty of Science, 
The University of Tokyo, 2-21-1 Osawa, Mitaka, Tokyo 181-8588, Japan.}} 
\altaffiltext{4}{{\footnotesize Department of Astronomy and Earth Sciences, 
Tokyo Gakugei University, 4-1-1 Nukui-kita, Koganei, Tokyo 184-8501, Japan.}}
\altaffiltext{5}{{\footnotesize Institute of Astrophysics and Planetary Science, 
Ibaraki University, 2-1-1 Bunkyo, Mito, Ibaraki 310-8512, Japan.}}

\begin{center}
{To appear in the Astrophysical Journal\\
(version: 18 October 1998)}
\end{center}

\begin{abstract}
We report the detection of a peculiar molecular cloud, CO 0.02--0.02, 
lying about 5$^\prime$ Galactic-east from the center of the Galaxy. 
$^{12}$CO images taken with NRO 45 m telescope showed that it is relatively compact ($ \sim\!3\times 4 $ pc$^2$) 
as well as having a very large velocity width ($\Delta V\!\geq\!100$ km s$^{-1}$).    
The cloud has a virial mass about one order of magnitude larger than the LTE mass, 
$9\times 10^4 M_{\sun}$, indicating the cloud is apparently gravitationally unbound.  

New observations with the JCMT 15 m and the NRO 45 m telescope show that CO 0.02--0.02 is 
very bright in the CO {\it J}=3--2, and in the HCN and HCO$^+$ {\it J}=1--0 lines. 
It appears that the environment may have an unusually high density and temperature, 
which may be related to the very broad CO line-width.  
We propose that CO 0.02--0.02 may have been accelerated, heated and compressed in 
a series of supernovae shocks which have occurred within the last $\mbox{(3--5)}\times 10^4$ years.  

\end{abstract}
\clearpage

\section{INTRODUCTION}
Molecular gas in the Galactic center region has a highly complex spatial distribution and kinematic structure, 
which has led to the identification of remarkable variety of peculiar features 
(Bania 1977, 1980, 1986; Bally et al. 1987, 1988; Uchida et al. 1992, 1996; Burton \& Liszt 1978, 1983, 1992; Oka et al. 1998b).   
It has long been known that giant molecular clouds in the region show large velocity widths, 
typically up to 5 times greater than those of the disk cloud population (Oka et al. 1998a). 
These results have led to the conclusion that the local environment has a higher pressure 
than typical of disc population material (Spergel \&\ Blitz 1992; Oka et al. 1998a). 

We recently made large-scale $^{12}$CO and $^{13}$CO images of the Galactic center region, 
mapping on a 34\arcsec\ grid with the $2\times 2$ focal-plane array receiver 
on the Nobeyama Radio Observatory (NRO) 45 m telescope (Oka et al. 1998b).  
In these data sets, those giant molecular clouds with large velocity widths do not always 
appear to be superpositions of clouds with `normal' velocity widths, 
but rather seem to consist of individual compact ($d\!\leq\!10$ pc) clouds having relatively large velocity widths ($\Delta V\!\geq\!30$ km s$^{-1}$).  
These compact clouds with large velocity widths are neither gravitationally bound, 
nor confined by the higher external pressure in the Galactic bulge, 
and thus may be transient features with lifetimes of $\sim\!10^5$ years.  
If they are of a transient nature, 
their frequency amongst the molecular cloud ensemble requires that they are being formed continuously.  

CO 0.02--0.02 is one such compact cloud having a large velocity width, 
which is centered at ($l, b)\!\simeq\!(+0.02$\arcdeg, $-0.02$\arcdeg), 
about 5\arcmin\ Galactic-east (we adopt this form of nomenclature to express relative direction 
in a galactic reference frame co-ordinate reference frame) from Sgr A$^*$ (see Fig.1{\it a}).   
This cloud stands out because of its broad line-width ($\Delta V\!\geq\!100$ \kms ) and its compact size ($\sim\!3\times 4 $ pc$^2$).   
The $^{12}$CO {\it J}=1--0 images show that this cloud is well isolated at velocities 
between $\VLSR\!=\!80$ \kms\ and 150 \kms .  
The velocity of the peak intensity toward CO 0.02--0.02, $\VLSR\!=\!90\,\kms $, 
is shifted by $\sim\!50\,\kms$ from the main ridge of intense CO emission from the Galactic center (Fig.1{\it b}).  
Even in our high resolution images, the cloud appears as a single entity, which distinctly differ from 
the Clump 1 (Bania, Stark, \&\ Heiligman, 1986) or the Clump 2 (Stark \&\ Bania 1986) .  
In addition, severely suffering the contamination of foreground gas, 
CO 0.02--0.02 seems to have a faint negative velocity extension to $\VLSR\!=\!-80$ \kms\ elongated towards the negative longitude.  

Here we report fully sampled CO {\it J}=3--2 observations, 
which are supplemented with HCN {\it J}=1--0, and HCO$^+$ {\it J}=1--0 observations, 
to clarify the spatial and velocity structure. 
Combined with the CO {\it J}=1--0 data, these data are then used to estimate the physical conditions in this unusual object. 
We adopt $D\!=\!8.5$ kpc as the distance to the Galactic center.

\section{OBSERVATIONS}
CO {\it J}=3--2 (345.795989 GHz) observations were obtained with the James Clerk Maxwell Telescope (JCMT) at Mauna Kea, Hawaii.  
An area covering  10\arcmin$\times$6\arcmin\ was mapped in the $^{12}$CO {\it J}=3--2 line on a 3.3\arcsec\ grid, 
which included CO 0.02--0.02 and the Galactic Center circumnuclear ring.  
A smaller region around CO 0.02--0.02 was also mapped in the $^{13}$CO {\it J}=3--2 line on the same grid spacing.  
The JCMT beamwidth was 14.3\arcsec\ full width at half maximum and the efficiency $\eta_{\rm fss}$ 
was 0.70 at 345 GHz (using the Facility Receiver B3i).  
Pointing errors were corrected by observing 345 GHz continuum emission from NGC 6334 I (nomenclature of McBreen et al. 1979).  
A digital autocorrelation spectrometer was used in its 920 MHz bandwidth mode, 
achieving an 800 \kms\ wide velocity coverage with 0.66 \kms\ resolution.  

The {\it J}=1--0 lines of HCN (88.631 GHz) and HCO$^+$ (89.188523 GHz) were measured 
with the NRO 45 m telescope using receivers S80 and S100.  
An area which included CO 0.02--0.02 and CO 0.13--0.13 (see Oka et al. 1998b for the nomenclature) 
was mapped on a grid with 17\arcsec\ sampling.  
The telescope has a FWHM beamwidth of 18\arcsec$\pm$1\arcsec\ at 86 GHz, 
and an $\eta_{\rm fss}$ of $\simeq\!0.58$ at 115 GHz (Oka et al 1998b).  
Pointing errors were corrected every two hours by observing the SiO maser emission (43 GHz) from VX Sgr.  
The pointing accuracy was better than 3\arcsec\ (rms) in both azimuth and elevation.  
Typical system noise temperatures of the receivers ranged from 400 to 600 K including atmospheric losses during the observations.  
We used wide-band acousto-optical spectrometers with an instantaneous bandwidth of 250 MHz and a frequency resolution of 250 kHz, 
which correspond to an 840 \kms\ wide velocity coverage with 0.84 \kms\ resolution.  
The observations were made by position switching to a clean reference position located at $(l, b)\!=\!(0\arcdeg, -2\arcdeg)$.  
Typical 60 seconds on-source integrations resulted in spectra having a noise level of $\sim\!0.15$  K rms.  
Spectra were also obtained of the H$^{13}$CN and H$^{13}$CO$^+$ {\it J}=1--0 isotopomers towards the center of CO 0.02--0.02, 
integrating for 22 minutes on-source.

\section{SPATIAL AND VELOCITY STRUCTURES}
Fig.2{\it a} shows the distribution of the CO {\it J}=3--2 emission 
integrated over the velocity range $-100$ to $+200$ \kms .   
This map delineates the cloud structure more clearly than does the {\it J}=1--0 data because of the better sampling 
--- however its distribution does confirm the general structure observed in CO {\it J}=1--0.  
CO 0.02--0.02 appears as a well defined, bright object in the integrated map, 
with an intense peak at ($l, b$)=($+0.013$, $-0.019$\arcdeg). CO 0.02--0.02, 
its `finger' like extension, and a chain of small clumps lie at the edge of 
an emission cavity ($\sim$ $3\times 2$ pc$^2$) which can be seen to the Galactic-southwest.  
Figs.2{\it b--c} show the distribution of HCN and HCO$^+$ {\it J}=1--0 emission 
integrated over the velocity range $-100$ to $+200$ \kms .  
CO 0.02--0.02 is also the brightest object in these integrated maps, 
and the southwestern emission cavity is still clearly seen.  
The other large cloud at lower latitude is the `$+50$ \kms  cloud', 
which dominates the HCN and HCO$^+$ maps in the velocity range between $+20$ and $+80$ \kms .  

Fig.3 shows the longitude-velocity map of CO{ \it J}=1--0 and HCN {\it J}=1--0 emissions around CO 0.02--0.02.  
In HCN {\it J}=1--0 emission, CO 0.02--0.02 appears with a well defined entity 
over a wide velocity range from $\VLSR\!\simeq\!40$ to $160\,\kms$.  
Its velocity width is larger than the other Galactic center clouds at least by a factor of 2.  
In addition, the clumps at $\VLSR\!\simeq\!10\,\kms$ and $-40$ \kms\ in the same longitude 
are likely negative-velocity extensions of CO 0.02--0.02, 
since they occupies nearly the same positions as the main body 
in the plane of the sky with small spatial extensions (see also Fig.4).  
However, the association of the clump at $(l, \VLSR)\!\simeq\!(-0.03\arcdeg, -70\,\kms)$ 
with CO 0.02-0.02 is controversial.  
Velocity channel maps showing the CO {\it J}=3--2 emission around CO 0.02--0.02 are reproduced in Fig.4.  
This emission has a strong peak at $\VLSR\!\simeq\!+90$ \kms .  
Despite the severe contamination of foreground gas at velocities close to $\VLSR\!\simeq\!-50,\,-30$, and 0 \kms , 
small emission clumps with negative velocity are associated with 
the spatial position of the intensity peak at $\VLSR\!=\!80$ to $100$ \kms .
Thus, the gas belong to CO 0.02--0.02 spans the velocity range from $\VLSR\!=\!-80$ to $+180$ \kms .     

Emission cavities associated with CO 0.02--0.02 also appear in velocity channel maps, 
at $(l, b, \VLSR)\!\simeq\!(0\arcdeg, -0.02\arcdeg, -50\,\kms)$, $(0.01\arcdeg, -0.03\arcdeg, 10\,\kms)$, 
$(0.025\arcdeg, -0.01\arcdeg, 30\,\kms)$, $(0\arcdeg, -0.025\arcdeg, 80\,\kms)$, and $(0.02\arcdeg, 0\arcdeg, 110\,\kms)$.  
The cavity at $(l, b, \VLSR)\!\simeq\!(0.02\arcdeg, 0\arcdeg, 110\,\kms)$ 
forms a shell-like structure with a diameter of 1 pc in the Galactic-northeastern portion of CO 0.02--0.02.  
The rounded head of CO 0.02--0.02 in the lower adjacent velocity channel ($\VLSR\!=\!+80$ to $+100$ \kms ) 
may be related to this small shell.  
Due to contamination by ambient emission, the lower velocity extent of the small shell remains uncertain, 
while its higher velocity emission merges with noise at $\VLSR\!\simeq\!+140$ \kms .   
The cavity at $(l, b, \VLSR)\!\simeq\!(0\arcdeg, -0.025\arcdeg, 80\,\kms)$ also forms 
a shell-like structure with a diameter of 2 pc.  
The southwestern emission cavity seen in the integrated intensity maps seems not be a coherent expanding shell, 
but the superpositions of several small cavities.  
All the molecular emission from CO 0.02--0.02 peaks at the position between the small shell-like structure 
at $(l, b, \VLSR)\!\simeq\!(0.02\arcdeg, 0\arcdeg, 110\,\kms)$ and the southwestern emission cavity.

\section{DISCUSSION}
\subsection{Physical Conditions}
Figs.5 and 6 show the CO, HCN, and HCO$^+$ spectra towards the center of CO 0.02--0.02.   
The line profiles clearly show the unusually large velocity width exhibited by this cloud.  
The line profiles are strongly peaked at $\VLSR\!\simeq\!+90$ \kms .  
The line intensity ratios at $\VLSR\!=\!+90$ \kms\ are listed in Table 1, 
and compared with those typical of molecular clouds in the galactic center and disk populations.  
The high values of the CO {\it J}=3--2/{\it J}=1--0 and HCN/CO {\it J}=1--0 ratios suggest that the lines 
are excited in a region having high temperature and high density relative to the normal disk cloud population.  

The high values of CO/$^{13}$CO {\it J}=1--0, CO/$^{13}$CO {\it J}=3--2, HCN/H$^{13}$CN {\it J}=1--0, 
and HCO$^+$/H$^{13}$CO$^+$ {\it J}=1--0 ratios imply relatively low optical depths towards this cloud; 
their optical depths are estimated to be; $\tau\!=$2--3 for CO and HCN lines; $\tau\!\ll\!1$ for HCO$^+$ line, 
assuming LTE conditions and adopting an abundance ratio [$^{12}$C]/[$^{13}$C]$\!=\!24$ (Langer \&\ Penzias 1990).   
An LVG analysis was made using the CO {\it J}=1--0 and {\it J}=3--2, and HCN {\it J}=1--0 line intensities, 
adopting [HCN]/[CO]=$2.5\times 10^{-8}/2.4\times 10^{-5}$ (Greaves \&\ Nyman 1996). 
This suggests that the material is excited in conditions where $n({\rm H}_2)\!\simeq\!10^{4.2}$ cm$^{-3}$, 
$T_{\rm k}\simeq 60$ K, and $N_{\rm CO}/dV \simeq 10^{17.1}$ cm$^{-2}$ (\kms )$^{-1}$.

\subsection{Total Mass}
The CO column density can be estimated from 
the integrated CO intensity by assuming that the rotational levels are thermalised at a temperature $T_{\rm ex}$.  
For the $^{12}$CO {\it J}=3--2 line, the general equation for estimating the column density is given by:
\begin{equation}
N_{\rm ^{12}CO} = 4.65\times 10^{12}\,T_{\rm ex}\,\mbox{exp}\left(\frac{33.2}{T_{\rm ex}}\right)\,
\int T_{\rm R}^*\mbox{(3--2)} dV\,\frac{\tau}{1-e^{-\tau}}
\end{equation}
where $\tau$ is the $^{12}$CO {\it J}=3--2 line optical depth, and $\tau/(1-e^{-\tau})$ is 
a correction factor which takes account of the optical depths.  
From this relationship, the optical depth towards the cloud center 
is estimated to be $2.7$, based on the observed $^{12}$CO/$^{13}$CO ratio.  
The excitation temperature between the $J\!=\!3$ and $2$ levels is estimated 
from the $^{12}$CO {\it J}=3--2 peak antenna temperature and the optical depth using the relationship:
\begin{equation}
T_{\rm R}^* = \frac{h\nu /k}{\mbox{exp}\left(h\nu /kT_{\rm ex}\right)-1} \left(1-e^{-\tau}\right), 
\end{equation}
to be $T_{\rm ex}\!=\!56$ K.  Taking the $^{12}$CO {\it J}=3--2 emission 
integrated over the velocity range $\VLSR\!=\!-100$ to $+200$ \kms ,  
adopting $T_{\rm ex}\!=\!56$ K and $[^{12}\mbox{CO}]/[\mbox{H}_2]\!=\!24\times 10^{-6}$ (Lis \&\ Goldsmith 1989, 1990; Langer \&\ Penzias 1990), 
and including the helium correction 1.36,
the total mass of CO 0.02--0.02 is then found to be $9.1\times 10^4$ M$_{\sun}$.  

An independent estimate of the mass can be obtained from the $^{13}$CO {\it J}=1--0 intensity, using the relationship:
\begin{equation}
N_{\rm ^{13}CO} = 4.58\times 10^{13}\,T_{\rm ex}\,\mbox{exp}\left(\frac{5.29}{T_{\rm ex}}\right)\,
\int T_{\rm R}^*\mbox{(1--0)} dV\,\frac{\tau}{1-e^{-\tau}}
\end{equation}
where $\tau$ is the $^{13}$CO {\it J}=1--0 line optical depth.  
Here the excitation temperature between the $J\!=\!1$ and $0$ levels 
is estimated from the CO {\it J}=1--0 peak antenna temperature, as 42 K.  
From the $^{13}$CO {\it J}=1--0 emission integrated over the velocity range 
$\VLSR\!=\!+50$ to $+150$ \kms , and adopting 
$T_{\rm ex}\!=\!42$ K and [$^{13}$CO]/[H$_2$]$\!=\!10^{-6}$ (Lis \&\ Goldsmith 1989, 1990), the mass $M = 2.6\times 10^5$ M$_{\sun}$, 
which is similar to the $^{12}$CO {\it J}=1--0 value, $M\!=\!3.1 \times 10^5$ M$_{\sun}$, 
but is larger than that derived from the CO {\it J}=3--2 intensity by a factor of 3.  

The discrepancies in the estimated masses and excitation temperatures 
may be due to the inapplicability of one-zone LTE analysis.  
The different transitions trace material in different degrees of excitation; for instance, 
the {\it J}=3--2 line of CO mainly trace highly excited gas, while the {\it J}=1--0 line traces moderately excited gas.  
We take $9\times 10^4$ M$_{\sun}$ as the LTE mass of CO 0.02--0.02, 
which is derived by the CO {\it J}=3--2 line intensity, 
since the CO {\it J}=1--0 intensities may suffer the contaminations of foreground and background gas with low excitation.  
From the LTE mass and the size of the cloud, $3\times 4\times 3$ pc$^3$, 
the average gas density is as high as $\langle n({\rm H}_2)\rangle\!\simeq\!5.6\times 10^4$ cm$^{-3}$, 
which is slightly higher than that estimated by the LVG analysis (\S 4.1).  
The LVG density and temperature may have suffered downward shifts 
by the contamination of gas with low excitation in the CO {\it J}=1--0 intensity.

\subsection{Energetics}
The size parameter and the velocity dispersion of CO 0.02--0.02 are measured 
from the $^{12}$CO {\it J}=3--2 data cube over the range, 
$l\!=\!-36$\arcsec\ to $+108$\arcsec, $b\!=\!-108$\arcsec\ to $+36$\arcsec, $\VLSR\!=\!-100$ to $+200$ \kms , 
to be $S\!=\!D {\rm tan}(\sqrt{\sigma_{l}\sigma_{b}})\!=\!1.6$ pc and $\sigma_{\rm V}\!=\!54$ \kms .    
If we take the velocity range $\VLSR\!=\!40$ to $200$ \kms (taking only the positive-velocity component), 
the velocity dispersion becomes $32$ \kms\ while the size parameter does not change.  
A virial theorem mass is determined by  
\begin{equation}
M_{\rm VT} = 3 f_p \frac{S \sigma_V^2}{G} , 
\end{equation}
where $f_p$ is a projection factor (Solomon et al. 1987; we took $f_p\!=\!2.9$).
The size and velocity dispersions of CO 0.02--0.02 give viral masses of 
$9.4\times 10^6\, M_{\sun}$ ($\sigma_{\rm V}\!=\!54$ \kms ) 
and $3.3\times 10^6\, M_{\sun}$ ($\sigma_{\rm V}\!=\!32$ \kms ), 
both of which are far larger than the LTE mass obtained in the preceding section.  
Thus, CO 0.02--0.02 is apparently gravitationally unbound.  

Given the average gas density of $\langle n({\rm H}_2)\rangle\!\simeq\!5.6\times 10^4$ cm$^{-3}$, 
a velocity dispersion of $32\,\kms$ corresponds to 
a kinetic pressure of $p/k\!\simeq\!5\times10^{9}$ cm$^{-3}$ K. 
This enormous kinetic pressure requires an unusual formation history, 
since the pressure from the Galactic bulge amounts only to 
$p/k\!\simeq\!5\times10^{6}$ cm$^{-3}$ K (Spergel \&\ Blitz 1992). 
The expansion time is $\mbox{(3--5)}\times 10^4$ years, 
which sets an upper limit to the age of CO 0.02--0.02.   
From the masses and velocity dispersions given above, 
the kinetic energy of CO 0.02--0.02 is $\mbox{(3--8)}\times 10^{51}$ erg.  
We consider two mechanisms for acceleration of the gas: 
supernovae explosions (type II, Ib), or Wolf-Rayet stellar winds.

Since a supernova loses large amount of its initial energy ($\sim\!10^{51}$ erg) through radiation, 
especially in high-density medium (e.g., Chevalier 1974), 
the kinetic energy estimated above requires an energy supply 
which might be available from several to several tens of supernovae.  
Multiple emission cavities in and around CO 0.02--0.02 do indeed suggest that 
acceleration could have been driven by a series of supernova explosions.  
The absence of a radio continuum source ($S_{\rm 20cm}\!\leq\!15$ mJy/beam; Yusef-Zadeh, Morris, \&\ Chance 1984) 
has been explained in terms of deficiency of relativistic electrons by synchrotron loss.  
In fact, given fields of 1 mG (e.g., Morris 1996), 
the synchrotron lifetime of a 0.55 GeV electron which able to contribute to radio emission at 20 cm, is $\sim\!1.5\times 10^4$ years.  
This is compatible with the notion that energy from the relativistic electrons produced in the supernova blast wave 
has already been radiated away by synchroton emission. 

A W-R star ($M\!\geq\!20$ M$_{\sun}$) injects kinetic energy into interstellar space 
as a stellar wind at a rate typically $\sim\!6\times 10^{44}$ erg yr$^{-1}$.  
If the energy source of CO 0.02--0.02 is W-R stellar wind(s), its kinetic energy and expansion time would 
require the presence of more than (110--450) W-R stars within the cloud.  
These massive stars would however also ionise the ambient gas, which would in turn emit thermal radio continuum.  
As an example; an O9 stars ($M\!=\!20$ M$_{\sun}$) on the ZAMS emit $10^{48.08}$ s$^{-1}$ Lyman continuum photons (Panagia 1979).    
The number of Lyman continuum photons $Q_{\rm Lyc}$ and the thermal radio flux $S_{\nu}$ are related by the equation 
\begin{equation}
Q_{\rm Lyc} = \frac{4.8\times 10^{48}}{\eta} T_{\rm e}^{-0.45} \nu^{0.1} S_{\nu} D^2
\end{equation}
(Mezger et al. 1979) , where $\eta$ is the dilution factor, $T_{\rm e}$ is the electron temperature, 
$\nu$ is the observed frequency in GHz, and $D$ is the distance to the source in kpc.  
If half of the Lyman continuum photons are taken up in ionising the ambient gas, 
and adopting $T_{\rm e}\!=\!10^4$ K, implies a radio flux $S_{\rm 10GHz}\!=\!\mbox{(10--40)}$ Jy.  
The observed flux of $S_{\rm 10GHz}\!\sim\!8$ Jy (Handa et al. 1987),   
and no enhancement exceeding 1 Jy toward CO 0.02--0.02, are incompatible with the above value.
Thus, radio continuum data do not favour acceleration of the material by the winds from a W-R star cluster.  

It seems most likely that CO 0.02--0.02 has been accelerated, 
heated and compressed by several tens of supernovae within the last $\mbox{(3--5)}\times 10^4$ years.  
Why this unusual object is in the vicinity of the Galactic nucleus ? 
What causes such a high supernova rate in this limited area ?
Although these queries still remain as mysteries, 
they may be related to the higher star formation activity in the recent past (Genzel 1997).  
A series of supernova explosions or a superposition of W-R stellar winds 
might produce a localized region of high pressure, 
especially in the region close to the center where many evolved stars have been found.  
The prevalence of such compact clouds with large velocity widths, 
as well as the boisterous molecular gas kinematics 
may also indicate that active star formation has occurred 
in the Galactic Center cloud ensemble 
as recently ago as several $\times 10^7$ years ago.

%

\acknowledgments
We thank JCMT staff and NRO staff for excellent support.   
We are also grateful to the anonymous referee for his or her comments 
and suggestions to improve the manuscript.  
T.O. is financially supported by the Special Postdoctoral Researchers Program of RIKEN.

\clearpage

\figcaption[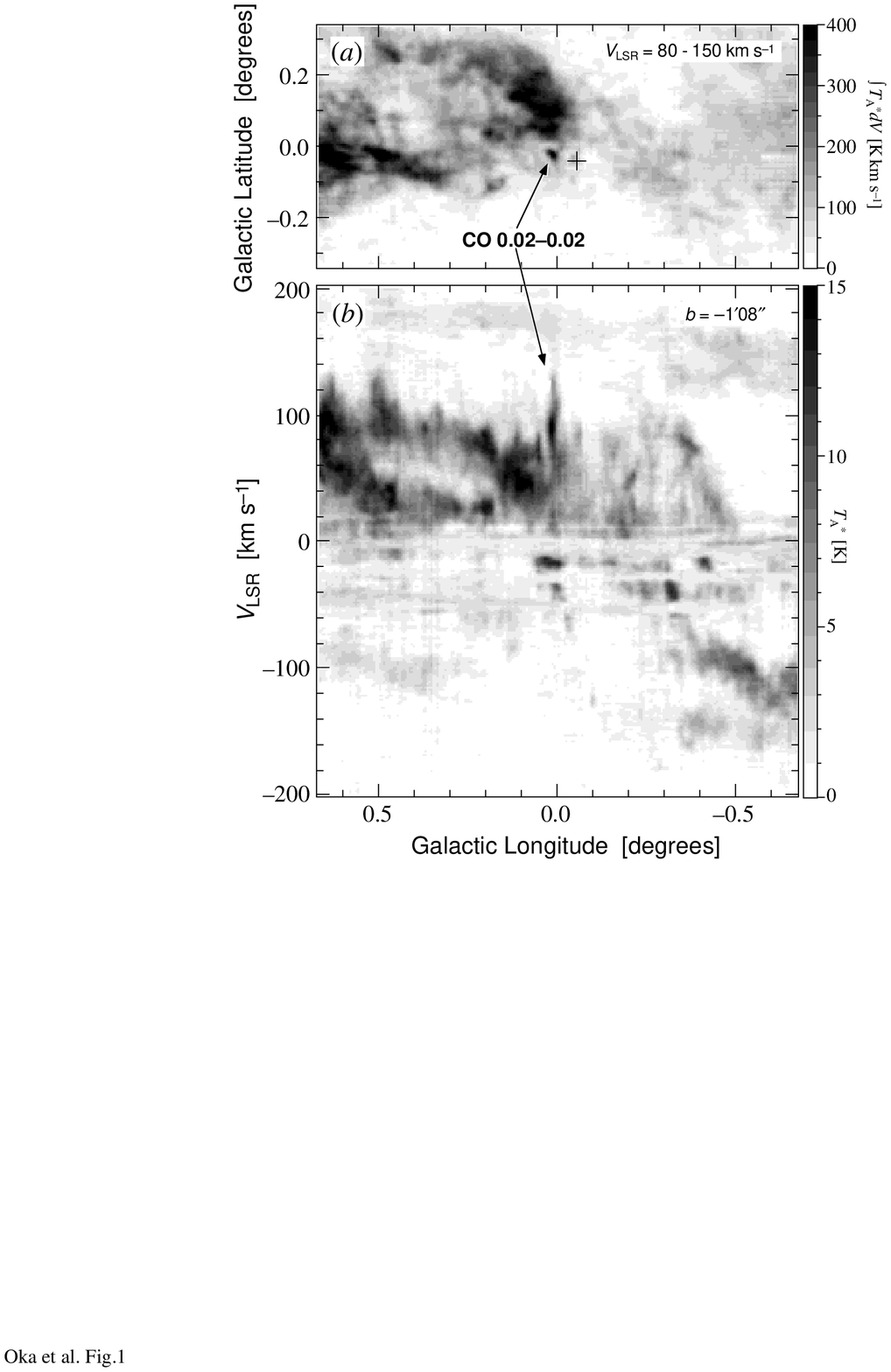]{({\it a}) A map of CO {\it J}=1--0 emission integrated over the velocity range 
$\VLSR= +80$ to $+150$ \kms .  A cross indicates the location of Sgr A$^*$ .  
({\it b}) A position-velocity map of CO {\it J}=1--0 emission at $b = -1$\arcmin 08\arcsec .  
\label{fig1}}

\figcaption[f2.eps]{({\it a}) A map of CO {\it J}=3--2 emission integrated over the velocity range 
$\VLSR= -100$ to $+200$ \kms .  
Contours are set at intervals of 200 K \kms .  White contours begin at 3000 K \kms .
({\it b}) A map of HCN {\it J}=1--0 emission around CO 0.02--0.02 integrated over the same velocity range as panel {\it a}).  
Contour intervals are 10 K \kms .  White contours begin at 250 K \kms .  
({\it c}) A map of HCO$^+$ {\it J}=1--0 emission around CO 0.02--0.02 integrated over the same velocity range as panel {\it a}).   
Contour intervals are 7.5 K \kms .  White contours begin at 135 K \kms .
\label{fig2}}

\figcaption[f3.eps]{({\it a}) A position-velocity map of CO {\it J}=1--0 emission at $b = -1$\arcmin 08\arcsec\ around CO 0.02--0.02.
Contours are set at intervals of 1 K .  White contours begin at 13 K .
({\it b}) A position-velocity map of HCN {\it J}=1--0 emission at $b = -1$\arcmin 08\arcsec\ around CO 0.02--0.02.
Contours are set at intervals of 0.2 K.  White contours begin at 2.6 K. 
\label{fig3}}

\figcaption[f4.eps]{Velocity channel maps of CO {\it J}=3--2 emission, each covering 
a velocity interval of 20 \kms , starting with $\VLSR\!=\!-100$ \kms\ and ending at $\VLSR\!=\!200$ \kms .  
Contour intervals are 40 K \kms .  White contours begin at 520 K \kms .
Crosses denote the position of the intensity peak at $\VLSR\!=\!800$ to $100$ \kms .  
\label{fig4}}

\figcaption[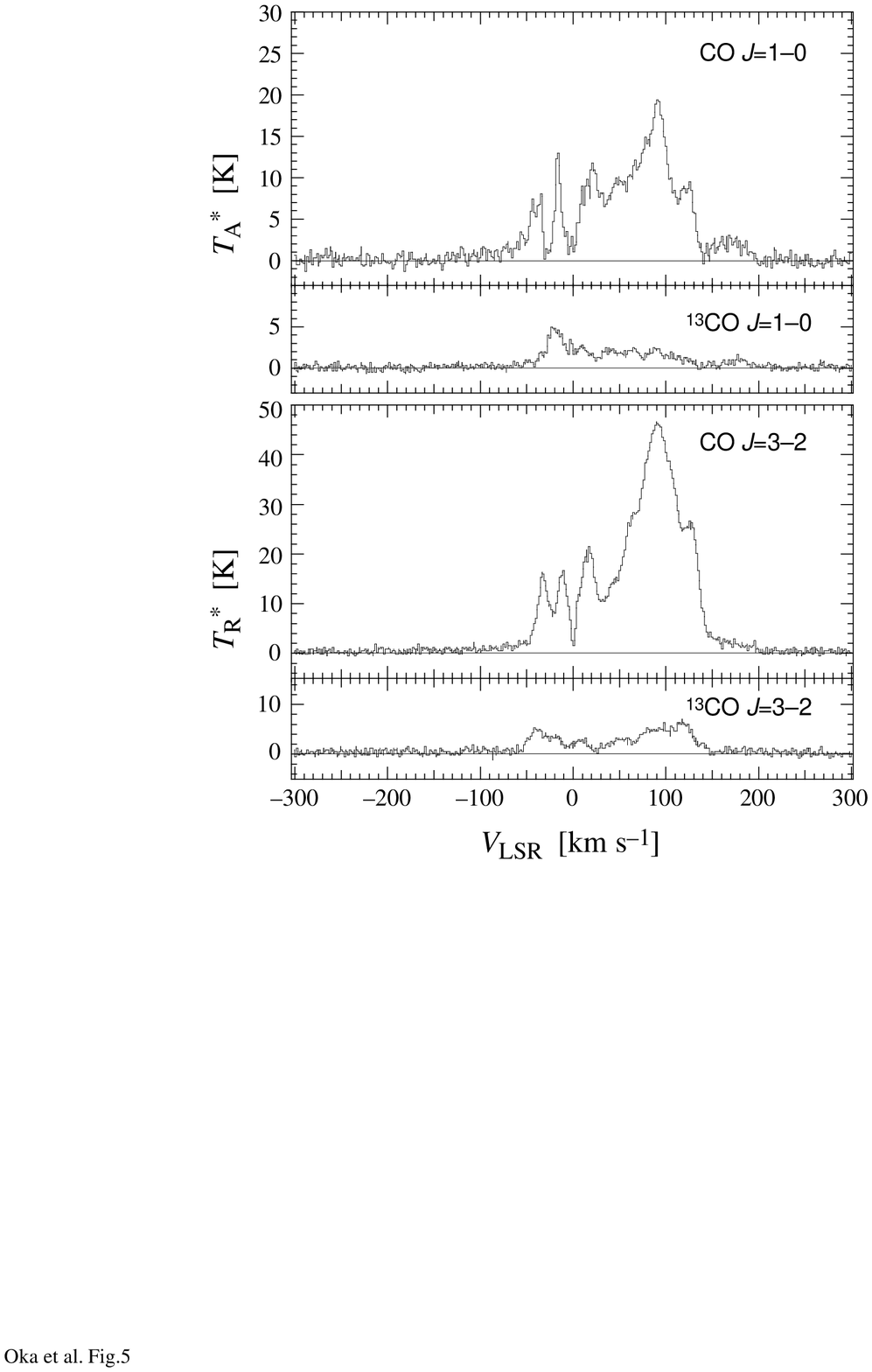]{Spectra of CO {\it J}=1--0 and $^{13}$CO {\it J}=1--0 toward $(l, b)\simeq (+0.019$\arcdeg, $-0.019$\arcdeg) 
taken with the NRO 45 m telescope,  
and spectra of CO {\it J}=3--2 and $^{13}$CO {\it J}=3--2 toward ($l, b$)=($+0.013$, $-0.019$\arcdeg) taken with JCMT.  
Weak emission humps in CO {\it J}=1--0 profiles at $\VLSR  = 150-200$ \kms\ is the positive velocity part of the expanding molecular ring, in which the high velocity wing of CO 0.02--0.02 lies buried.  
\label{fig5}}

\figcaption[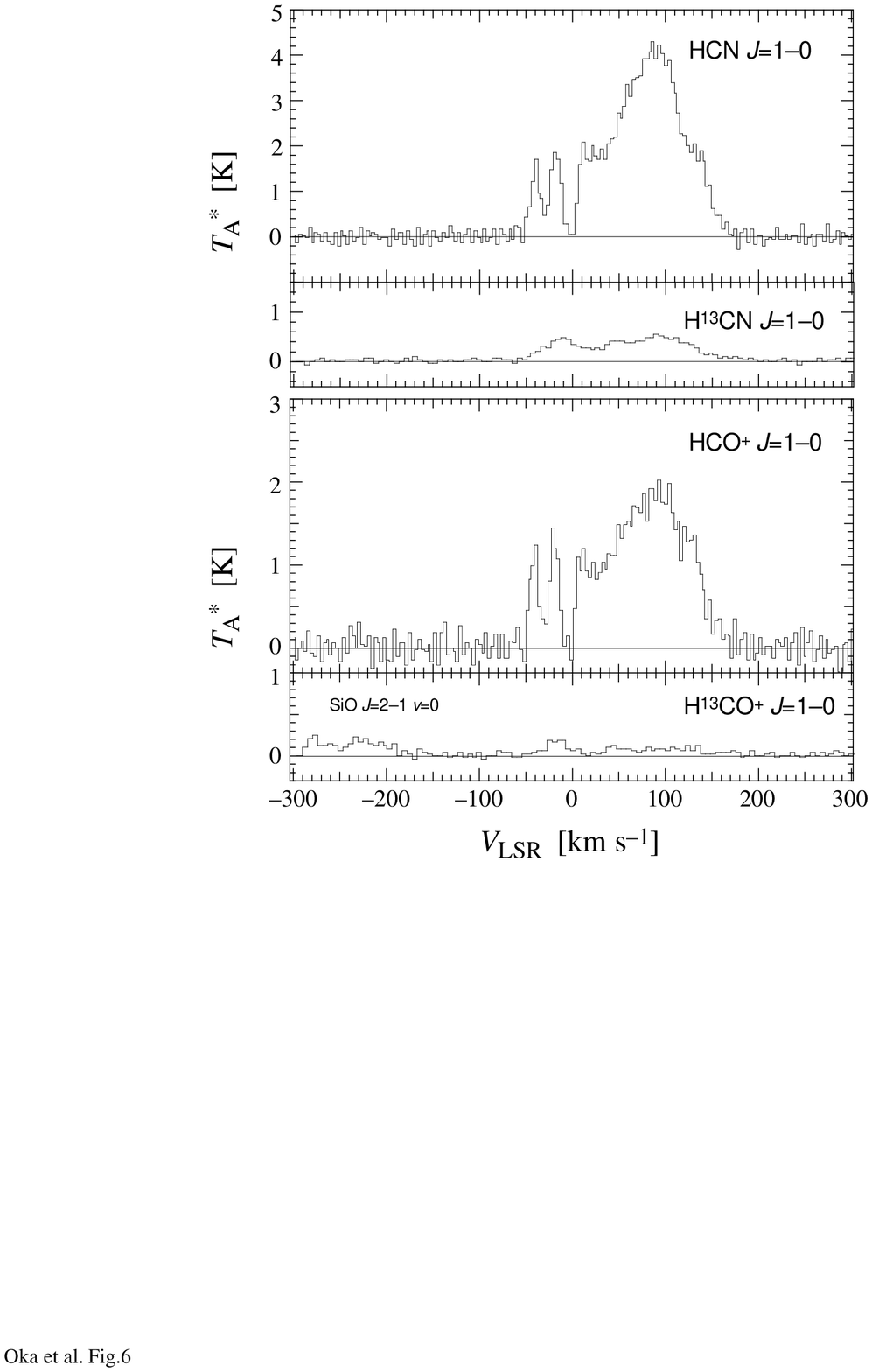]{Spectra of HCN {\it J}=1--0, H$^{13}$CN {\it J}=1--0, HCO$^+$ {\it J}=1--0 , 
and H$^{13}$CO$^+$ {\it J}=1--0 toward $(l, b)\simeq (+0.019$\arcdeg, $-0.019$\arcdeg) taken with the NRO 45 m telescope.  
\label{fig6}}

\begin{deluxetable}{llll}
\tablewidth{30pc}
\tablecaption{Line Intensity Ratios}
\tablehead{
\colhead{line}  & \colhead{CO 0.02--0.02} & \colhead{GC clouds} & \colhead{disk clouds}
}
\startdata
CO/$^{13}$CO {\it J}=1--0	& 8	& 5.2\tablenotemark{a} & $5-8$\tablenotemark{b} \nl
CO/$^{13}$CO {\it J}=3--2	& 11	&  \nodata &  \nodata \nl
CO {\it J}=3--2/{\it J}=1--0	& 1.3	&  0.4\tablenotemark{c}  &  0.3\tablenotemark{c} \nl
HCN/CO {\it J}=1--0		&  0.2	&   $0.08$\tablenotemark{d,e}  &  $0.01-0.05$\tablenotemark{e} \nl
HCN/H$^{13}$CN {\it J}=1--0	& 8  &  \nodata  &   \nodata \nl
HCO$^+$/H$^{13}$CO$^+$ {\it J}=1--0	& $> 25$  &  \nodata  &  $1-6$\tablenotemark{f} \nl
\tablenotetext{a}{From large-scale surveys (Oka et al. 1998b).}
\tablenotetext{b}{From Galactic plane surveys (Liszt  1993).}
\tablenotetext{c}{From the CfA survey (Dame et al. 1987) and the COBE data (Bennett et al. 1994).}
\tablenotetext{d}{From large-scale surveys (Jackson et al. 1996).}
\tablenotetext{e}{From Galactic plane surveys (Helfer \&\ Blitz 1997).}
\tablenotetext{f}{For nearby clouds (Gu\'elin, Langer, \& Wilson 1982).}
\enddata
\end{deluxetable}


\begin{thebibliography}{}
\bibitem[]{}Bally, J., Stark, A.  A., Wilson, R.  W., \& Henkel, C.  1987, ApJS, 65, 13
\bibitem[]{}Bally, J., Stark, A.  A., Wilson, R.  W., \& Henkel, C.  1988, ApJ, 324, 223  
\bibitem[]{}Bania, T. M.  1977, ApJ, 216, 381   
\bibitem[]{}Bania, T. M.  1980, ApJ, 242, 95  
\bibitem[]{}Bania, T. M.  1986, ApJ, 308, 868 
\bibitem[]{}Bania, T.ÊM., Stark,ÊA.ÊA., Heiligman,ÊG.ÊM.  1986, ApJ, 307, 350  
\bibitem[]{}Bennett, C. L., Fixsen, D.J., Hinshaw, G., Mather, J. C., Moseley, S. H., Wright, E. L., Eplee, Jr., R. E.,  Gales, J., Hewagama, T., Isaacman, R. B., Shafer, R. A., \&\ Turpie, K.  1994, ApJ, 434, 587  
\bibitem[]{}Burton, W.  B., \& Liszt, H.  S.  1978, ApJ, 225, 815  
\bibitem[]{}Burton, W.  B., \& Liszt, H.  S.  1983, ApJS, 52, 63  
\bibitem[]{}Burton, W. B., \& Liszt, H. S.  1992, A\&AS, 95, 9  
\bibitem[]{}Chevalier, R. A.  1974, ApJ, 188, 501  
\bibitem[]{}Dame, T. M.,  Ungerechts, H., Cohen, R. S., de Geus, E. J., Grenier, I. A., May, J., Murphy, D. C., Nyman, L. -\AA., \&\ Thaddeus, P.  1987, ApJ, 322, 706  
\bibitem[]{}Genzel, R.  1998, in IAU Symp.184, The Central Regions of the Galaxy and Galaxies. ed. Y. Sofue, in press  
\bibitem[]{}Greaves, J. S., \&\ Nyman, L. -\AA.  1996, A\&A, 305, 950  
\bibitem[]{}Gu\'elin, M., Langer, W. D., \&\ Wilson, R. W.  1982, A\&A, 107, 107
\bibitem[]{}Handa, T. , Sofue, Y. , Nakai, N. , Hirabayashi, H. , \&\ Inoue, M.  1987, PASJ, 39, 709  
\bibitem[]{}Helfer, T. T., \&\ Blitz, L  1997, 478, 233  
\bibitem[]{}Jackson, J. M.,  Heyer, M. H., Paglione, T. A. D., \&\ Bolatto, A. D.  1996, ApJ, 456. L91  
\bibitem[]{}Langer, W. D., \& Penzias, A. A.  1990, ApJ, 357, 477  
\bibitem[]{}Lis, D. C., \& Goldsmith, P. F.  1989, ApJ, 337, 704  
\bibitem[]{}Lis, D. C., \& Goldsmith, P. F.  1990, ApJ, 356, 195  
\bibitem[]{}Liszt, H. S.  1993, ApJ, 411, 720  
\bibitem[]{}McBreen, B., Fazio, G. G., Stier, M., \& Wright, E. L.  1979, ApJ, 232, L183
\bibitem[]{}Mezger, P. G., Pankonin, V., Schmidt-Burgk, J., Thun, C., \&\ Wink, J.  1979, A\&A, 80, L3
\bibitem[]{}Morris, M.  1996, in IAU Symp. 169, Unsolved Problems of the Milky Way, eds. L. Blitz\& P. Teuben, (Dordrecht: Kluwer), 247    
\bibitem[]{}Oka, T., Hasegawa, T., Hayashi, M., Handa, T., \& Sakamoto, S.  1998a, ApJ, 493, 730  
\bibitem[]{}Oka, T., Hasegawa, T., Sato, F., Tsuboi, M., \& Miyazaki, A.  1998b,  ApJS in press
\bibitem[]{}Panagia, N.  1979, AJ, 78, 929  
\bibitem[]{}Solomon, P. M., Rivolo, A. R., Barrett, J., \&\ Yahil, A.  1987, ApJ, 319, 730  
\bibitem[]{}Spergel, N., \&\ Blitz, L.  1992, Nature, 357, 665  
\bibitem[]{}Stark, A. A., \&\ Bania, T. M. 1986, ApJ, 306, L17  
\bibitem[]{}Uchida, K. I., Morris, M., Bally, J., Pound, M., \&\ Yusef-Zadeh, F.  1992, ApJ, 398, 128  
\bibitem[]{}Uchida, K. I., Morris, M., Serabyn, E., \&\ G\"usten, R.  1996, ApJ, 462, 768  
\bibitem[]{}Yusef-Zadeh, F., Morris, M., \& Chance, D.  1984, Nature, 310, 557  
\end{thebibliography}
\end{document}